\documentclass[12pt,letterpaper]{article}

\usepackage[T1]{fontenc}
\usepackage[utf8]{inputenc}
\usepackage[margin=1in]{geometry}
\usepackage{amsmath,amssymb}
\usepackage{booktabs}
\newcommand{\E}{\mathbb{E}}
\renewcommand{\P}{\mathbb{P}}
\usepackage{pgfplots}
\pgfplotsset{compat=1.18}
\usepgfplotslibrary{groupplots}
\definecolor{ltcteal}{HTML}{00798C}
\definecolor{ltcrose}{HTML}{D1495B}
\definecolor{ltcmist}{HTML}{C8CFD6}
\usepackage[round,authoryear]{natbib}
\usepackage{authblk}
\usepackage{microtype}
\usepackage{parskip}
\usepackage{xurl}
\usepackage[colorlinks=true,allcolors=blue]{hyperref}

\title{Challenge or Pass? A Dynamic Policy for MLB's\\
Automated Ball-Strike System}
\author[1]{Jonathan Pipping-Gam\'on}
\author[2]{Zach Sissman}
\author[3]{Lila Dodson}
\author[1]{Abraham J. Wyner}
\affil[1]{Department of Statistics \& Data Science, University of Pennsylvania}
\affil[2]{Community School of Naples}
\affil[3]{San Francisco University High School}
\date{}

\hypersetup{
  pdftitle={Challenge or Pass? A Dynamic Policy for MLB's Automated Ball-Strike System},
  pdfauthor={Jonathan Pipping-Gam\'on, Zach Sissman, Lila Dodson, Abraham J. Wyner}
}

\begin{document}
\maketitle

\begin{abstract}
Under Major League Baseball's automated ball-strike (ABS) challenge system, players must weigh the value of overturning a call against the risk of losing a challenge they may need later.
We develop a policy that values an overturn through its effect on expected runs, accounting for the count, outs, and runners on base.
Using pitch tracking and game records, we reconstruct challenge opportunities, including calls that players let stand.
A Bayesian model represents players' imperfect impressions of pitch location and uses them to estimate overturn probabilities, with uncertainty calibrated from historical challenge decisions.
Dynamic programming compares the expected gain with the future cost of losing a challenge, accounting for the stage of the game, the number of challenges remaining, and opportunities on both offense and defense.
We develop the policy using games before the 2026 All-Star break and evaluate it without refitting on 983 games played afterward.
Along the recorded sequence of opportunities, the policy recovers 0.536 modeled correction runs per game across both teams, compared with 0.452 under observed decisions, an improvement of 18.5\%.
The gain is 0.084 runs per game, with a 95\% bootstrap interval of $[0.065, 0.103]$.
The policy makes more challenges and recovers more value per successful review despite a lower overturn rate.
These results show why overturn rate alone can be a misleading measure of challenge strategy.
\end{abstract}

\section{Introduction}
\label{sec:introduction}

\subsection{What makes a challenge valuable?}
\label{sec:challenge-value}

Major League Baseball's automated ball-strike (ABS) challenge system, introduced in 2026, gives players the opportunity to immediately review an umpire's ball or strike call. A successful challenge by a batter reverses a called strike---improving the count or even earning a walk with ball four---while a successful catcher's or pitcher's challenge does the opposite, converting a called ball into a strike. But challenging a call also carries risk: if the umpire's initial call was right, the team loses one of its two challenges, and with it the chance to correct a potentially more consequential call later in the game \citep{mlb2025absrules}. Players are therefore forced to balance the potential benefit of an overturn with the cost of losing a challenge.

To illustrate this, consider two calls made in the first inning, each with two outs and two challenges still available. With a full count and runners on first and second, Royals outfielder Starling Marte took a called third strike against the Dodgers. Public pitch tracking places the pitch outside the ABS zone, but Marte declined to challenge, ending the inning and stranding two runners. In another game, Reds' infielder Nathaniel Lowe successfully challenged a first-pitch strike against the Guardians with the bases empty. His review put him ahead 1--0 instead of behind 0--1, but there were still two outs and nobody on base.

These game situations suggest that Marte had far more to gain from a potential overturn than Lowe, and run expectancy is one way to describe exactly how much more \citep{lindsey1963strategies}. Using 2023--2025 scoring data, we estimate how many runs a team could expect to score from a particular base-out-count state, comparing these expectations under the original and corrected calls. For Marte, loading the bases with two outs is worth 0.782 expected runs more than ending the inning, while for Lowe, improving the count from 0--1 to 1--0 is worth 0.032.

\subsection{When is a challenge worth the risk?}
\label{sec:present-and-future}

But the potential benefit of overturning a call is only one consideration when deciding whether to challenge. Players must also judge how likely the umpire was to be wrong, knowing that an unsuccessful challenge may cost their team an opportunity to correct a future call. A relatively small gain like Lowe's may still be worth challenging when an overturn is nearly certain, while a consequential call like Marte's may not justify the risk if the chances of success are slim. The decision therefore depends not only on the call at hand, but also on the value of preserving a challenge for later.

That value changes as the game unfolds. Losing both challenges early leaves a team unable to contest another missed call for the rest of regulation, while losing the second in the ninth leaves far fewer opportunities for that loss to matter. And because successful challenges are retained, a single challenge can potentially correct multiple calls over the course of a game. Putting a value on a lost challenge is more difficult, however, because we do not yet know which calls a team will face or how it will respond. We therefore estimate how many challenge opportunities remain and compare the runs a team can expect to gain from correcting future calls with its current challenges and with one fewer. By expressing both the immediate benefit of an overturn and the future cost of an unsuccessful challenge in expected runs, we can determine how likely an overturn must be to justify the risk.

\subsection{Our approach and contribution}

The problem of deciding when to use a limited number of challenges is not unique to baseball. In professional tennis, \citet{abramitzky2012linecalls} study how players should balance the potential benefit of correcting a line call against the risk of losing a challenge they may need later. Applying this framework to baseball requires understanding both the likelihood that a called ball or strike was incorrect and the consequences of correcting it. Existing research provides a foundation for each. Pitch-framing models examine how pitch location, count, and the players involved influence umpire calls \citep{deshpande2017pitchframing}, while plate-discipline models evaluate decisions at the plate through their effects on expected runs \citep{yee2024platediscipline}.

Recent work has also begun to examine ABS challenge decisions directly. \citet{martinez2026abs} formulates the allocation of challenges as a Markov decision process and learns a policy in a simulated environment built from 2024 Statcast data. \citet{lee2026algorithmic} instead examine umpire calls and how players actually use their challenges under the 2026 system, including how those decisions relate to information about the pitch and the number of challenges remaining.

Our approach begins with an important distinction between what we can measure after a pitch and what a player knows when deciding whether to challenge. Pitch-tracking data allow us to estimate where a ball crossed the plate, but players must judge its location without that information. We model their judgments as noisy signals of pitch location, using them alongside the umpire's call and the count to estimate the probability of an overturn. Historical challenge decisions help us calibrate this uncertainty, although they cannot tell us how much comes from errors in judging a pitch and how much comes from errors in deciding whether to challenge. We then use historical sequences of called pitches to estimate the future cost of losing a challenge, accounting for opportunities on both offense and defense. Comparing that cost with the immediate run value of an overturn allows us to determine how likely a challenge must be to succeed before it is worth the risk.

We develop this policy using games from before the 2026 All-Star break and evaluate it on 983 games played afterward, including calls that players chose not to challenge. Because overturning a call could change the pitches and situations that follow, we evaluate the policy against the recorded sequence of play rather than attempting to reconstruct how each game would have unfolded differently. Under this evaluation, the policy recovers an estimated 0.536 expected runs per game, compared with 0.452 under observed decisions, an improvement of approximately 19\%. It achieves this improvement despite a lower overturn rate, suggesting that challenging the right calls matters more than maximizing the percentage of successful reviews.

\label{sec:contribution}
\section{Data}
\label{sec:data}

\subsection{Calls and challenge opportunities}
\label{sec:decision-record}

The challenges players actually make tell only part of the story.
Every eligible called strike presents a potential challenge opportunity for the batter, while every called ball presents one for the catcher or pitcher.
We therefore construct a pitch-by-pitch record of calls that meet ABS review eligibility, including those left unchallenged and those occurring after a team has exhausted its challenges.
These later calls matter because a different sequence of earlier decisions might have left the team with a challenge available.

We build this record using MLB's live game feed and Statcast data from Baseball Savant \citep{mlb2026statsapi,mlb2026savant}.
MLB's live game feed records the sequence of play, umpire calls, and challenge outcomes, while Statcast provides pitch locations and the count, outs, and runners before each pitch. We match pitches across the two sources by game, plate appearance, and pitch number, using review outcomes to recover the umpire's original call when a pitch is recorded under its corrected ruling. We also track each team's remaining challenges throughout the game, accounting for the restoration of one challenge when a team enters extra innings with none left \citep{mlb2025absrules}. This allows us to distinguish the calls players could actually challenge from those they might have been able to challenge under a different strategy.

\subsection{Which calls could have been overturned?}
\label{sec:observed-and-perceived-location}

For pitches that were challenged, we observe whether ABS upheld or overturned the umpire's call \citep{mlb2026abs}.
Most calls are never reviewed, however, leaving us without an official ruling on whether they were correct.
To evaluate the decisions players made---and the challenges they passed up---we need to assess those calls as well.

We do so by comparing each pitch's tracked location with the batter-specific ABS strike zone, accounting for the ball's radius and the corners of the zone \citep{mlb2025absrules,mlb2026savant}.
A called strike is correctable when the ball misses the zone entirely, while a called ball is correctable when any portion of it crosses the zone.
This reconstruction allows us to assess calls regardless of whether they were challenged, although our classifications remain subject to error in the publicly available tracking data.

To describe how far a pitch was from the strike-zone boundary, we define a signed margin $M$, measured in inches, with its direction determined by who could challenge the call. For a batter challenging a called strike, a positive margin means the pitch was outside the zone. For a catcher or pitcher challenging a called ball, a positive margin means the pitch was inside the zone. In either case, $M>0$ indicates a correctable call. A pitch on the boundary counts as a strike, so a called ball at $M=0$ is also correctable. Appendix~\ref{app:geometry} details how margins are calculated.

\subsection{Development and evaluation samples}
\label{sec:samples}

To measure the value of overturning a call, we use Statcast data from 2023--2025 to estimate the expected runs remaining in an inning across different count, out, and baserunner situations. We then develop the challenge policy using the first half of the 2026 season and evaluate its recommendations on games played after the All-Star break, allowing us to test the policy on a fresh sample of games.

The development sample covers 1,442 ABS-enabled games played from March 25 through July 12, containing 211,596 eligible tracked called pitches and 6,042 observed challenges. The evaluation sample covers 983 games played from July 16 through the end of the season, containing 143,126 eligible tracked called pitches and 4,516 observed challenges, of which 2,468 were successful. We evaluate the policy on these games without refitting. Both samples exclude two games that lacked ABS infrastructure.

\section{A run-value challenge policy}
\label{sec:policy}

\subsection{The value of correcting a call}
\label{sec:correction-gain}

A successful challenge can change the count, put a batter on base, or end an inning. We compare what those outcomes are worth by their effect on expected runs \citep{lindsey1963strategies}.
Let $\widehat{\mathrm{RE}}(x)$ be the estimated runs remaining in the half-inning from base-out-count state $x$.

We estimate this quantity for the 288 combinations of baserunners $b$, outs $o$, and count $c$, using pitches from 2023--2025.
To give sparsely observed counts support from a larger sample, we shrink each estimate toward the mean for the same baserunners and outs across all counts.
Let $Y_j$ be the runs scored from pitch $j$ through the end of its half-inning, $n_{boc}$ the number of observations in state $(b,o,c)$, and $\bar Y_{bo}$ the mean across counts for that base--out state.
The estimated runs remaining from state $(b,o,c)$ are
\begin{equation}
 \widehat{\mathrm{RE}}(b,o,c)
 =\frac{\displaystyle\sum_{j:x_j=(b,o,c)}Y_j+\lambda_{\mathrm{RE}}\bar Y_{bo}}
        {n_{boc}+\lambda_{\mathrm{RE}}}.
 \label{eq:re-shrinkage}
\end{equation}
The pooled mean contributes the equivalent of $\lambda_{\mathrm{RE}}$ additional pitches, so it has more influence in states with fewer observations.
Predictions for later historical seasons determine how strongly these estimates borrow across counts, with $\lambda_{\mathrm{RE}}=100$ minimizing mean squared error.
Estimation and validation details appear in Appendices~\ref{app:run-expectancy} and~\ref{app:pooling-validation}.

With these estimates, we compare the expected runs under the ball and strike rulings for each pitch.
For pitch $i$ and ruling $a$, let $R_i(a)$ be the runs that would score immediately, including any run forced home by ball four, and let $x_i(a)$ be the resulting base-out-count state.
The gain from correcting the call is
\begin{equation}
 G_i = R_i(\mathrm{ball})+\widehat{\mathrm{RE}}\{x_i(\mathrm{ball})\}
      -R_i(\mathrm{strike})-\widehat{\mathrm{RE}}\{x_i(\mathrm{strike})\}.
 \label{eq:correction-gain}
\end{equation}
For a batter challenging a called strike, $G_i$ measures the increase in expected runs if the call becomes a ball.
For a catcher or pitcher challenging a called ball, the same difference measures the reduction in expected runs allowed if the call becomes a strike.

To construct these two outcomes, we keep the recorded runner movements, outs, and runs on the pitch, then apply the consequences of each ruling.
Ball four advances the batter and any forced runners, while strike three adds an out.
If that is the third out, the half-inning ends and remaining run expectancy is zero.
This calculation tells us what a successful challenge would be worth.
Whether that potential gain justifies a challenge, however, also depends on how likely the call is to be overturned.

\subsection{From perceived location to overturn probability}
\label{sec:overturn-probability}

The umpire's call and the count provide information about where a pitch is likely to have crossed the plate.
For a batter considering a challenge, we start with the margins of called strikes in the same count.
For a catcher or pitcher, the corresponding distribution comes from called balls.
We use these distributions as priors for the pitch's margin.

For each role $r\in\{o,d\}$ (offense or defense) and pre-pitch count $c$, we estimate this distribution from the development sample of challengeable pitches.
We group margins into 0.01-inch bins and shrink the frequencies for each count toward the overall distribution for that role, allowing counts with fewer observations to draw on a larger sample of pitches.
Let $n_{rc}(m)$ be the number of pitches in the bin centered at $m$ for role $r$ and count $c$, let $n_{rc}$ be their total across bins, and let $p_r(m)$ be the empirical bin frequency for that role across all counts.
The resulting prior mass is
\begin{equation}
 p_{rc}(m)=\frac{n_{rc}(m)+\alpha_r p_r(m)}{n_{rc}+\alpha_r}.
 \label{eq:margin-prior}
\end{equation}
The parameter $\alpha_r$ determines how strongly the count-specific distribution is pulled toward the role-wide distribution.
Its value is chosen separately for offense and defense by predictive performance in held-out development games (Appendix~\ref{app:pooling-validation}).

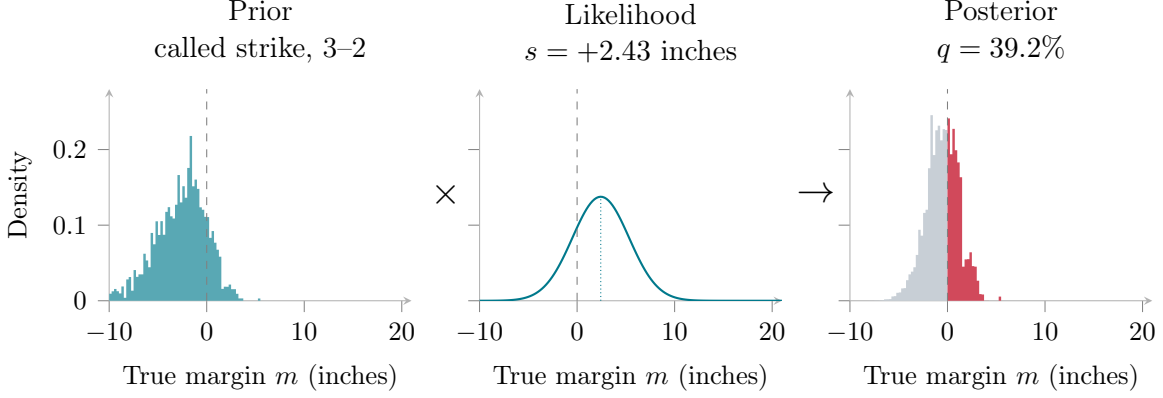
\begin{figure}[htbp]
\centering
\begin{tikzpicture}
\begin{groupplot}[
 group style={group size=3 by 1,horizontal sep=0.9cm},
 scale only axis,width=4cm,height=2.8cm,
 xmin=-10,xmax=21,ymin=0,ymax=0.28,
 xtick={-10,0,10,20},ytick={0,0.1,0.2},
 scaled ticks=false,
 xlabel={True margin $m$ (inches)},
 tick label style={font=\footnotesize},label style={font=\footnotesize},
 title style={font=\small,align=center},
 axis lines=left,axis line style={gray!60},tick align=outside,
]
\nextgroupplot[title={Prior\\called strike, 3--2},ylabel={Density}]
\addplot[ybar interval,fill=ltcteal!65,draw=none] coordinates {
(-10,0.0093) (-9.75,0.0125) (-9.5,0.0156) (-9.25,0.0126)
(-9,0.0097) (-8.75,0.0189) (-8.5,0.0039) (-8.25,0.0282)
(-8,0.0314) (-7.75,0.0133) (-7.5,0.0406) (-7.25,0.0316)
(-7,0.0349) (-6.75,0.0349) (-6.5,0.0622) (-6.25,0.0533)
(-6,0.0443) (-5.75,0.0897) (-5.5,0.0747) (-5.25,0.1051)
(-5,0.087) (-4.75,0.1053) (-4.5,0.0873) (-4.25,0.1176)
(-4,0.1238) (-3.75,0.1117) (-3.5,0.1088) (-3.25,0.127)
(-3,0.1663) (-2.75,0.1301) (-2.5,0.1515) (-2.25,0.1363)
(-2,0.1757) (-1.75,0.218) (-1.5,0.1514) (-1.25,0.1603)
(-1,0.148) (-0.75,0.1236) (-0.5,0.1204) (-0.25,0.111)
(0,0.1107) (0.25,0.0832) (0.5,0.0922) (0.75,0.0768)
(1,0.0675) (1.25,0.0583) (1.5,0.0157) (1.75,0.0186)
(2,0.0185) (2.25,0.0214) (2.5,0.0153) (2.75,0.0153)
(3,0.0092) (3.25,0.0031) (3.5,0.0031) (3.75,0)
(4,0) (4.25,0) (4.5,0) (4.75,0)
(5,0) (5.25,0.003) (5.5,0) (5.75,0)
(6,0) (6.25,0) (6.5,0) (6.75,0)
(7,0) (7.25,0) (7.5,0) (7.75,0)
(8,0) (8.25,0) (8.5,0) (8.75,0)
(9,0) (9.25,0) (9.5,0) (9.75,0)
(10,0) (10.25,0) (10.5,0) (10.75,0)
(11,0) (11.25,0) (11.5,0) (11.75,0)
(12,0) (12.25,0) (12.5,0) (12.75,0)
(13,0) (13.25,0) (13.5,0) (13.75,0)
(14,0) (14.25,0) (14.5,0) (14.75,0)
(15,0) (15.25,0) (15.5,0) (15.75,0)
(16,0) (16.25,0) (16.5,0) (16.75,0)
(17,0) (17.25,0) (17.5,0) (17.75,0)
(18,0) (18.25,0) (18.5,0) (18.75,0)
(19,0) (19.25,0) (19.5,0) (19.75,0)
(20,0) (20.25,0) (20.5,0) (20.75,0)
(21,0)
};
\draw[gray,dashed] (axis cs:0,0) -- (axis cs:0,0.28);
\nextgroupplot[title={Likelihood\\$s=+2.43$ inches},yticklabels=\empty]
\addplot[ltcteal,thick,domain=-10:21,samples=151]
 {exp(-((x-2.42535623969)^2)/(2*2.89985356706^2))/(2.89985356706*sqrt(2*pi))};
\draw[gray,dashed] (axis cs:0,0) -- (axis cs:0,0.28);
\draw[ltcteal,densely dotted] (axis cs:2.42535623969,0)
 -- (axis cs:2.42535623969,0.137573250226);
\nextgroupplot[title={Posterior\\$q=39.2\%$},yticklabels=\empty]
\addplot[ybar interval,fill=ltcmist,draw=none] coordinates {
(-10,0) (-9.75,0) (-9.5,0) (-9.25,0)
(-9,0) (-8.75,0) (-8.5,0) (-8.25,0.0001)
(-8,0.0002) (-7.75,0.0001) (-7.5,0.0004) (-7.25,0.0004)
(-7,0.0006) (-6.75,0.0008) (-6.5,0.0019) (-6.25,0.002)
(-6,0.0022) (-5.75,0.0057) (-5.5,0.0061) (-5.25,0.0105)
(-5,0.011) (-4.75,0.0163) (-4.5,0.0167) (-4.25,0.0272)
(-4,0.0349) (-3.75,0.0385) (-3.5,0.0443) (-3.25,0.0606)
(-3,0.0939) (-2.75,0.0861) (-2.5,0.1162) (-2.25,0.1204)
(-2,0.1753) (-1.75,0.2455) (-1.5,0.1927) (-1.25,0.2255)
(-1,0.2316) (-0.75,0.2122) (-0.5,0.227) (-0.25,0.2259)
(0,0) (0.25,0) (0.5,0) (0.75,0)
(1,0) (1.25,0) (1.5,0) (1.75,0)
(2,0) (2.25,0) (2.5,0) (2.75,0)
(3,0) (3.25,0) (3.5,0) (3.75,0)
(4,0) (4.25,0) (4.5,0) (4.75,0)
(5,0) (5.25,0) (5.5,0) (5.75,0)
(6,0) (6.25,0) (6.5,0) (6.75,0)
(7,0) (7.25,0) (7.5,0) (7.75,0)
(8,0) (8.25,0) (8.5,0) (8.75,0)
(9,0) (9.25,0) (9.5,0) (9.75,0)
(10,0) (10.25,0) (10.5,0) (10.75,0)
(11,0) (11.25,0) (11.5,0) (11.75,0)
(12,0) (12.25,0) (12.5,0) (12.75,0)
(13,0) (13.25,0) (13.5,0) (13.75,0)
(14,0) (14.25,0) (14.5,0) (14.75,0)
(15,0) (15.25,0) (15.5,0) (15.75,0)
(16,0) (16.25,0) (16.5,0) (16.75,0)
(17,0) (17.25,0) (17.5,0) (17.75,0)
(18,0) (18.25,0) (18.5,0) (18.75,0)
(19,0) (19.25,0) (19.5,0) (19.75,0)
(20,0) (20.25,0) (20.5,0) (20.75,0)
(21,0)
};
\addplot[ybar interval,fill=ltcrose,draw=none] coordinates {
(-10,0) (-9.75,0) (-9.5,0) (-9.25,0)
(-9,0) (-8.75,0) (-8.5,0) (-8.25,0)
(-8,0) (-7.75,0) (-7.5,0) (-7.25,0)
(-7,0) (-6.75,0) (-6.5,0) (-6.25,0)
(-6,0) (-5.75,0) (-5.5,0) (-5.25,0)
(-5,0) (-4.75,0) (-4.5,0) (-4.25,0)
(-4,0) (-3.75,0) (-3.5,0) (-3.25,0)
(-3,0) (-2.75,0) (-2.5,0) (-2.25,0)
(-2,0) (-1.75,0) (-1.5,0) (-1.25,0)
(-1,0) (-0.75,0) (-0.5,0) (-0.25,0)
(0,0.2412) (0.25,0.1938) (0.5,0.2274) (0.75,0.1991)
(1,0.183) (1.25,0.1634) (1.5,0.0451) (1.75,0.0547)
(2,0.055) (2.25,0.0641) (2.5,0.0458) (2.75,0.0451)
(3,0.0265) (3.25,0.0088) (3.5,0.0083) (3.75,0.0001)
(4,0.0001) (4.25,0) (4.5,0) (4.75,0)
(5,0) (5.25,0.0055) (5.5,0) (5.75,0)
(6,0) (6.25,0) (6.5,0) (6.75,0)
(7,0) (7.25,0) (7.5,0) (7.75,0)
(8,0) (8.25,0) (8.5,0) (8.75,0)
(9,0) (9.25,0) (9.5,0) (9.75,0)
(10,0) (10.25,0) (10.5,0) (10.75,0)
(11,0) (11.25,0) (11.5,0) (11.75,0)
(12,0) (12.25,0) (12.5,0) (12.75,0)
(13,0) (13.25,0) (13.5,0) (13.75,0)
(14,0) (14.25,0) (14.5,0) (14.75,0)
(15,0) (15.25,0) (15.5,0) (15.75,0)
(16,0) (16.25,0) (16.5,0) (16.75,0)
(17,0) (17.25,0) (17.5,0) (17.75,0)
(18,0) (18.25,0) (18.5,0) (18.75,0)
(19,0) (19.25,0) (19.5,0) (19.75,0)
(20,0) (20.25,0) (20.5,0) (20.75,0)
(21,0)
};
\draw[gray,dashed] (axis cs:0,0) -- (axis cs:0,0.28);
\end{groupplot}
\path (group c1r1.east) -- node[font=\large] {$\times$} (group c2r1.west);
\path (group c2r1.east) -- node[font=\large] {$\to$} (group c3r1.west);
\end{tikzpicture}
\caption{For an illustrative impression at Marte's tracked margin,
$s=+2.43$ inches, the full-count prior and Gaussian likelihood yield a
posterior with 39.2\% of its mass on correctable calls. The primary
allocation uses a sensory standard deviation of 2.90 inches
(Section~\ref{sec:behavioral-calibration}).}
\label{fig:marte-confidence}
\end{figure}

We combine this prior with the player's own impression of where the pitch crossed the plate. Because that impression may be imprecise, we represent it as a perceived margin $S_i$, equal to the pitch's margin $M_i$ plus sensory error:
\begin{equation}
 S_i=M_i+\varepsilon_{s,i},\qquad
 \varepsilon_{s,i}\mid M_i,r,c\sim N(0,\sigma_{s,r}^2).
 \label{eq:perceived-margin}
\end{equation}
The standard deviation $\sigma_{s,r}$ describes uncertainty about pitch location in inches and can differ between offense and defense.
For $\sigma_{s,r}>0$, Bayes' rule combines the prior with the player's impression to give the probability $\P(M>0\mid S=s,r,c)$ that the call is correctable \citep{gelman2013bayesian}:
\begin{equation}
 q_{rc}(s;\sigma_{s,r})
 =\frac{\sum_{m>0}p_{rc}(m)\phi_{\sigma_{s,r}}(s-m)}
        {\sum_m p_{rc}(m)\phi_{\sigma_{s,r}}(s-m)},
 \label{eq:overturn-posterior}
\end{equation}
Here $\phi_\sigma$ is the normal density with mean zero and standard deviation $\sigma$.
Each possible margin is weighted by its prior probability and how likely it is to produce the player's impression. The numerator sums these weights over margins for which the call is correctable, while the denominator sums them over all possible margins.
For pitch $i$, we denote this posterior probability of a successful challenge by $q_i=q_{r_i c_i}(S_i;\sigma_{s,r_i})$.
Figure~\ref{fig:marte-confidence} shows the update for an illustrative impression of Marte's full-count pitch, using the sensory error scale specified below.

\subsection{What historical decisions tell us about uncertainty}
\label{sec:behavioral-calibration}

Observed challenge decisions reflect both where players thought a pitch was and how they acted on that judgment.
We model players as comparing their perceived margin with a cutoff $t_r(X_i)$, measured in inches, that depends on the game context $X_i$.
A higher cutoff means that the player needs to be more convinced the call was wrong before challenging.
To allow decisions to vary even for the same perceived margin and game context, we introduce an action error $\varepsilon_{a,i}$ with standard deviation $\sigma_{a,r}$.
Given the game context, we assume that sensory and action errors are normally distributed with mean zero and are independent of each other and of the pitch's true margin.
For pitches on which a challenge was available, let $A_i^{\mathrm{hist}}$ equal one if the player challenged and zero otherwise. We model this decision as
\begin{equation}
\begin{aligned}
 A_i^{\mathrm{hist}}
   &=\mathbf1\{M_i+\varepsilon_{s,i}+\varepsilon_{a,i}>t_r(X_i)\},\\
 \P(A_i^{\mathrm{hist}}=1\mid M_i,X_i,r)
   &=\Phi\!\left(\frac{M_i-t_r(X_i)}{w_r}\right),
 \qquad w_r^2=\sigma_{s,r}^2+\sigma_{a,r}^2,
\end{aligned}
 \label{eq:historical-choice}
\end{equation}
where $\Phi$ is the standard normal distribution function.
Holding game context fixed, a larger $w_r$ means that challenge probability rises more gradually as the pitch's true margin increases.

We fit separate probit models for offense and defense using calls within three inches of the strike-zone boundary.
The models adjust for the value of a correction, count, inning, score, outs, and remaining challenges.
They also account for pitch family, handedness matchup, and participant and team differences \citep{wood2017gam}.
For each role, the fitted coefficient on margin, $b_r$, gives the combined error width as $w_r=1/b_r$ inches.
The estimated widths are $2.900$ inches for offense and $1.823$ inches for defense.
Appendix~\ref{app:error-calibration} specifies the calibration model.

These widths combine sensory and action error, and observed decisions cannot identify the two components separately.
Our primary analysis assigns the full fitted scale to sensory error, taking $\sigma_{s,r}=w_r$ and $\sigma_{a,r}=0$.
We also assume that players follow the recommended policy given their impressions.
Section~\ref{sec:benchmarks-sensitivity} examines how the results change when the same total variance is divided differently between sensory and action error.

\subsection{Future value and the Bellman-optimal policy}
\label{sec:bellman-policy}

A failed review costs the team one of its remaining challenges, reducing its ability to correct calls later in the game.
We value that loss by comparing the runs a team could expect to recover from those later calls with its current challenges and with one fewer.
Let $C_k(z)$ be the expected run value of corrected pitches after the current opportunity, with $k$ challenges remaining and optimal decisions from then on.
Here $z$ records the half-inning, the number of outs, and whether the team is batting or fielding.
The team draws from the same pool of $k$ remaining challenges in both roles.
The cost of losing one challenge can therefore be expressed as
\begin{equation}
 L_k(z)=C_k(z)-C_{k-1}(z),\qquad k=1,2.
 \label{eq:inventory-loss}
\end{equation}
If the call is overturned, the team gains $G$ in expected runs and keeps its challenge; otherwise, it loses $L_k(z)$ in future value.
With overturn probability $q$, the decision therefore weighs the expected gain $qG$ from correcting the current call against the expected future cost $(1-q)L_k(z)$.
This is the tradeoff studied by \citet{abramitzky2012linecalls} for tennis challenges, expressed here in expected runs.
When a challenge is available, the policy follows the rule
\begin{equation}
 \text{challenge if }qG>(1-q)L_k(z).
 \label{eq:challenge-rule}
\end{equation}
A larger potential correction can therefore justify challenging with less confidence, while more valuable future opportunities favor preserving the challenge.
Because posterior overturn probability increases with perceived margin, this same rule can also be written as $S>t$, with the cutoff determined by the role, count, gain, and cost $L_k(z)$ of losing a challenge.

To estimate what opportunities remain, we follow each team's offensive and defensive calls through the development games.
These sequences show the state in which the next opportunity occurs and when opportunities end.
Calls that lack ABS eligibility or usable tracking remain in these sequences as mandatory passes, contributing no correction value while retaining information about how play progresses.
Regulation innings remain distinct, while corresponding situations in extra innings share a state.

These games also tell us how often different counts and correction values occur together.
We choose how much to pool this joint distribution and the next-call distribution across innings by comparing mean modeled correction runs recovered in held-out development games.
This selects $\lambda_F=0$, so both distributions are estimated from calls in the same state, without borrowing across innings.
We combine those frequencies with the margin priors and the model of players' impressions to estimate how often future calls offer different combinations of overturn probability $q$ and correction value $G$.
For a given role and count, this calculation assumes the same margin prior and sensory uncertainty across correction values and game stages.
The model assumes that $z$ alone determines the distribution of later opportunities, regardless of whether the team challenges the current call or lets it stand.
Further details appear in Appendices~\ref{app:future-opportunities}--\ref{app:pooling-validation}.

The value of preserving a challenge depends in turn on how the team will use it when those later calls arrive.
Let $V_k(z)$ be the expected correction value from the current opportunity onward, before its $q$ and $G$ are known.
Passing preserves the team's remaining challenges and their future value $C_k(z)$.
Challenging changes that value by $qG-(1-q)L_k(z)$, the expected gain from correcting the current call minus the expected cost of losing a challenge.
For each possible opportunity, the team chooses the action with greater expected value.
Averaging that value over the opportunities that can arise in state $z$ gives the Bellman equation
\begin{equation}
 V_k(z)=C_k(z)+\E\!\left[
   \max\{0,qG-(1-q)L_k(z)\}\mid z\right],\qquad k=1,2.
 \label{eq:bellman}
\end{equation}
When challenging offers no positive net benefit, including when $G\leq0$, the policy passes and the term inside the expectation is zero.
With no challenges remaining, the team must also pass, so $V_0(z)=C_0(z)$.
That future value can still be positive because entering an extra inning may restore a challenge.

To express that future value, let $\widehat P_z(z',e)$ be the estimated probability that the next call occurs in state $z'$, with $e=1$ if reaching it involves entering a later extra inning and $e=0$ otherwise.
If $k$ challenges remain after the current decision, the team has $g(k,e)$ available at the next call: one when $k=0$ and $e=1$, and $k$ otherwise.
Averaging the value from each possible next call gives
\begin{equation}
 C_k(z)=\sum_{z'}\sum_{e=0}^{1}
 \widehat P_z(z',e)\,V_{g(k,e)}(z').
 \label{eq:appendix-continuation}
\end{equation}
The probability that no further call occurs contributes zero future value.

We solve these coupled equations by value iteration \citep{puterman1994mdp}, starting with both $C$ and $V$ set to zero.
Each iteration updates $C$ from the current values of $V$, then updates $V$ with the Bellman equation, until the values stabilize.

\section{Policy evaluation}
\label{sec:evaluation}

\subsection{Evaluating decisions through whole games}
\label{sec:whole-game-evaluation}

For the primary analysis, we apply the fitted probability model and Bellman cost function to the test games without refitting.
We use each pitch's tracked margin and the sensory error model to describe the impressions a player might have of its location.
Given a particular impression and number of remaining challenges, the policy uses the posterior overturn probability to recommend either challenging or passing.
Averaging over the possible impressions gives the probability of choosing to challenge at each number of remaining challenges.
We compute this average directly from the sensory error distribution.
Conditional on the recorded pitch information, sensory errors are independent across pitches.

Starting with two challenges per team, we follow the recorded calls on both offense and defense.
Different impressions on earlier pitches can lead to different decisions, so a team may reach a later pitch with zero, one, or two challenges remaining.
We track the probability of each possibility and use those probabilities to weight the policy's chance of challenging at each level.
The reconstructed margin determines whether a challenge would succeed.
Failed challenges reduce the number remaining, while passes and successful challenges preserve it.
Entering an extra inning with no challenges restores one.

A correctable call contributes its correction value $G_i$ only when the policy challenges it, so its expected contribution is $G_i$ times the resulting probability of a challenge.
Correct calls contribute no correction value.
We sum these expected gains across both teams to report value per game, keeping the recorded pitches and baseball states fixed.
The result measures the value of correcting calls along those recorded opportunities rather than changes in final game scores.

\subsection{Benchmarks, sensitivity, and uncertainty}
\label{sec:benchmarks-sensitivity}

We compare the policy with observed decisions and a perfect-execution
oracle that knows each pitch's exact location. On the same recorded
opportunities, the oracle challenges every correctable call with $G_i>0$.
All its reviews succeed, so it retains its challenges and captures all
positive correction value available on this record.

Historical challenge decisions identify the combined error variance, but do not tell us how much comes from perceiving the pitch and how much comes from acting on that perception.
We therefore examine different divisions of this variance between sensory and action error, allowing any action error to continue affecting decisions under the recommended policy:
\begin{equation}
 \sigma_{s,r}=\kappa_r w_r,\qquad
 \sigma_{a,r}=w_r\sqrt{1-\kappa_r^2}.
 \label{eq:error-allocation}
\end{equation}
Here $\kappa_r\in[0,1]$, and $\kappa_r^2$ is the share of the total variance $w_r^2$ assigned to sensory error.
At $\kappa_r=0$, perception is exact and the entire fitted variance is assigned to action error.
The oracle removes both sources of error.

With a challenge available and $G>0$, a player using margin cutoff $t$ challenges when $S+\varepsilon_a>t$, or equivalently when $M+\varepsilon_s+\varepsilon_a>t$.
Changing the division between sensory and action error leaves their sum Gaussian with mean zero and variance $w_r^2$.
For the same true margin, game context, and cutoff, the probability of choosing to challenge is therefore unchanged.
It has the same probit form as the historical-choice model in Equation~\eqref{eq:historical-choice}, with $t$ replacing the historical threshold.

With the other fitted inputs held fixed, each cutoff therefore produces the same expected correction gains and the same probabilities of retaining or losing a challenge across these divisions of the error variance.
Optimizing within this cutoff family gives the same optimal policy value, so the cutoffs from our primary $\kappa_r=1$ analysis also attain that value under the other divisions.

The gain over observed decisions remains between 0.0828 and 0.0839 modeled correction runs per game when each pooling strength is varied in turn to neighboring grid values.
All comparisons use the selected run-expectancy table to value corrections (Appendix~\ref{app:pooling-validation}).

We form 95\% percentile intervals from 5,000 bootstrap resamples of the
983 test games \citep{efron1994bootstrap}. Each resample keeps both teams and all decisions within a
game together, using the same games to compare the policy with observed
decisions. The fitted inputs and policy remain fixed, so these intervals
capture variation across test games and exclude uncertainty from model
estimation.

\section{Results}
\label{sec:results}

\subsection{Recovered run value}
\label{sec:recovered-value}

Across the 983 test games, the Bellman-optimal policy recovers 0.536 modeled
correction runs per game, compared with 0.452 under observed decisions.
The gain is 0.084 runs per game (95\% interval $[0.065, 0.103]$), an
increase of 18.5\%. Both offense and defense contribute to this improvement,
with offensive opportunities accounting for nearly three-quarters of the
total gain (Table~\ref{tab:recovered-value}).
The policy captures 39.8\% of the correction value available to the oracle.

\begin{table}[htbp]
\centering
\caption{Total modeled runs recovered across both teams per game in the
983 test games. Gain is the policy's improvement over observed decisions,
with paired bootstrap 95\% intervals in brackets.}
\label{tab:recovered-value}
\begin{tabular}{lrrrr}
\toprule
Role & Observed & Bellman-optimal & Oracle & Gain (95\% interval) \\
\midrule
Offense & 0.181 & 0.243 & 0.700 & $0.061\;[0.049, 0.074]$ \\
Defense & 0.271 & 0.293 & 0.648 & $0.022\;[0.009, 0.036]$ \\
\midrule
Combined & 0.452 & 0.536 & 1.347 & $0.084\;[0.065, 0.103]$ \\
\bottomrule
\end{tabular}
\end{table}

\subsection{How the policy uses challenges}
\label{sec:challenge-use}

The policy makes more challenges: 6.14 per game across both teams,
compared with 4.59 observed. Its modeled overturn rate is 43.8\%, compared with 54.7\% under observed
decisions, but the additional attempts produce more successful reviews. Those
successes also recover more value on average, 0.199 runs compared with
0.180 under observed decisions. The improvement thus combines more
corrections with more valuable corrections, while accepting additional
failed reviews.

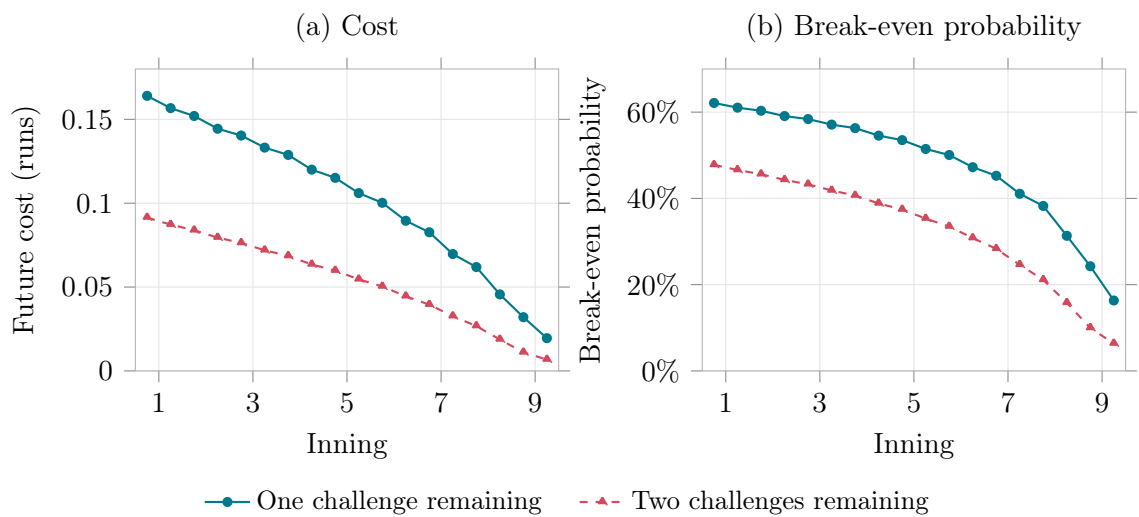
\begin{figure}[htbp]
\centering
\begin{tikzpicture}
\begin{groupplot}[
 group style={group size=2 by 1,horizontal sep=1.9cm},
 scale only axis,width=5.6cm,height=4cm,scaled y ticks=false,
 xmin=0.5,xmax=18.5,ymin=0,ymax=0.18,
 xtick={1.5,5.5,9.5,13.5,17.5},xticklabels={1,3,5,7,9},
 ytick={0,0.05,0.1,0.15},yticklabels={0,0.05,0.1,0.15},
 xlabel={Inning},ylabel={Future cost (runs)},
 tick label style={font=\small},label style={font=\small},
 title style={font=\small},
 grid=major,major grid style={ltcmist!55},
 axis line style={gray!60},tick align=outside,
 legend style={font=\footnotesize,draw=none,legend columns=2,
   /tikz/every even column/.append style={column sep=1em}},
]
\nextgroupplot[title={(a) Cost},legend to name=bellmaninventorylegend]
\addplot[ltcteal,thick,mark=*,mark size=1.5pt] coordinates {
 (1,0.163996322) (2,0.156672734) (3,0.151980207)
 (4,0.144390300) (5,0.140333453) (6,0.133130997)
 (7,0.128771263) (8,0.120004369) (9,0.115101360)
 (10,0.106012007) (11,0.100231143) (12,0.089511923)
 (13,0.082649799) (14,0.069682278) (15,0.061927670)
 (16,0.045629597) (17,0.032040374) (18,0.019514788)
};
\addlegendentry{One challenge remaining}
\addplot[ltcrose,thick,dashed,mark=triangle*,mark size=1.7pt] coordinates {
 (1,0.091439942) (2,0.087209640) (3,0.083831945)
 (4,0.079487259) (5,0.076266993) (6,0.071926236)
 (7,0.068521920) (8,0.063481784) (9,0.059806777)
 (10,0.054609418) (11,0.050338828) (12,0.044517740)
 (13,0.039490836) (14,0.032640102) (15,0.026748071)
 (16,0.018752144) (17,0.011063575) (18,0.006785626)
};
\addlegendentry{Two challenges remaining}
\nextgroupplot[title={(b) Break-even probability},
 ylabel={Break-even probability},ymax=0.7,ytick={0,0.2,0.4,0.6},
 yticklabels={0\%,20\%,40\%,60\%}]
\addplot[ltcteal,thick,mark=*,mark size=1.5pt] coordinates {
 (1,0.621206844) (2,0.610398820) (3,0.603143432)
 (4,0.590818458) (5,0.583911441) (6,0.571056612)
 (7,0.562882161) (8,0.545463572) (9,0.535102893)
 (10,0.514591400) (11,0.500577190) (12,0.472328714)
 (13,0.452504189) (14,0.410663263) (15,0.382440319)
 (16,0.313326397) (17,0.242655889) (18,0.163283459)
};
\addplot[ltcrose,thick,dashed,mark=triangle*,mark size=1.7pt] coordinates {
 (1,0.477642968) (2,0.465839473) (3,0.456024904)
 (4,0.442857391) (5,0.432678813) (6,0.418355206)
 (7,0.406605385) (8,0.388311057) (9,0.374244309)
 (10,0.353208872) (11,0.334835842) (12,0.308043429)
 (13,0.283107029) (14,0.246080192) (15,0.211033355)
 (16,0.157909943) (17,0.099614797) (18,0.063544379)
};
\end{groupplot}
\end{tikzpicture}
\par\smallskip
\pgfplotslegendfromname{bellmaninventorylegend}
\caption{Future correction value lost on failure (left) and break-even
overturn probabilities for a 0.10-run correction (right), with one or two
challenges remaining. Points follow a home team through regulation, with
defense in each top half followed by offense in the bottom half, evaluated
at calls with no outs. All costs include future opportunities in both roles.}
\label{fig:bellman-inventory}
\end{figure}

The cost of a failed challenge depends on the opportunities still ahead.
Figure~\ref{fig:bellman-inventory} follows a home team through regulation at calls with no outs, showing both the expected future correction value lost on failure and the overturn probability needed to justify a challenge for a 0.10-run gain.
Losing the last challenge costs more than losing one of two, so the policy requires greater confidence when only one remains.
As regulation progresses, fewer opportunities remain and both costs fall, allowing the team to pursue the same gain with less certainty.

For example, a home batter facing a 0.10-run correction opportunity with
no outs in the first inning needs an overturn probability above about
47\% with two challenges remaining, compared with 61\% when only one
remains. The higher confidence requirement reflects the cost of losing
future opportunities, even though the current call offers the same gain.

\subsection{Returning to the opening examples}
\label{sec:opening-examples}

For Marte and Lowe, losing one of two challenges carries a similar cost: 0.090 runs of expected future correction value for Marte's opportunity and 0.085 for Lowe's.
An overturn is worth much more to Marte, however, with a correction gain of 0.782 runs compared with 0.032 for Lowe.
To justify a challenge, the posterior probability of an overturn must therefore exceed 10.3\% for Marte but 72.6\% for Lowe.
Although both calls occur early in the game, the chance to preserve Marte's rally justifies challenging with substantially less certainty.

We then use each pitch's tracked location and the sensory error model to determine how often a player's impression would meet that requirement.
With two challenges available, the policy recommends challenging for 95.2\% of the modeled impressions of Marte's pitch and 0.8\% of those of Lowe's.
The policy therefore usually contests the valuable call left unchallenged and passes on the small correction that Lowe successfully obtained.
A sufficiently convincing impression could still justify Lowe's challenge, but its successful outcome alone does not tell us whether his confidence met the requirement.

\section{Discussion}
\label{sec:discussion}

\subsection{Implications for challenge strategy}
\label{sec:strategy-implications}

The policy recovers more modeled run value by changing which calls a team challenges while holding the model’s assumed uncertainty about pitch location fixed.
Its recommendations depend on both the value of correcting the current call and the opportunities the team might lose after a failed challenge.
Judging a strategy by its overturn rate alone overlooks calls the team lets stand and how much each correction is worth, which can favor caution even when a lower-confidence challenge has positive expected value.

A team's decisions at the plate also affect its ability to challenge calls in the field, because both roles draw from the same pool of challenges.
Preserving a challenge on offense can leave it available for a valuable defensive opportunity later in the game.
The policy can help teams review these decisions and develop guidance for players, with the confidence required to challenge reflecting both the current correction's value and the opportunities still ahead.

\subsection{Limitations and next steps}
\label{sec:limitations}

Our reported gains measure the value of correcting calls along the recorded sequence of pitches and game states.
An overturned call could change the pitches and situations that follow, so these gains do not estimate changes in final scores.
Our assessment also depends on the reconstructed strike zone and tracked pitch locations; errors in either can change which calls we classify as correctable.

Players' impressions are unobserved, so we model how they perceive pitches and act on those perceptions.
We assume Gaussian errors and a common error scale within each role, and condition the prior distributions on the umpire's call and the count.
The policy's optimality is conditional on this model, and individual differences or additional private information could change its recommendations.

Historical choices cannot tell us how much of the fitted uncertainty comes from perception and how much comes from action.
In Section~\ref{sec:benchmarks-sensitivity}, dividing that uncertainty differently leaves the optimal policy value unchanged within the specified margin-cutoff family.
This result relies on independent, mean-zero Gaussian errors, with the total variance and other fitted inputs held fixed.
Direct measures of player confidence could help assess the perception model and support calibration for individual players.

Bootstrap intervals capture variation across test games with fitted inputs held
fixed. Resampling the estimation data and refitting the inputs and policy
would also reflect estimation uncertainty.

Extending the objective to win probability would value a correction by its effect on the team's chance of winning.
Under that objective, the same gain in expected runs could be worth more or less depending on the score and stage of the game.

\section{Reproducibility}
\label{sec:reproducibility}

The analysis data and source code are available at
\url{https://github.com/WhartonSABI/abs}. The repository includes
instructions for reproducing the analysis.

\clearpage
\bibliographystyle{plainnat}
\bibliography{references}

\begin{thebibliography}{15}
\providecommand{\natexlab}[1]{#1}
\providecommand{\url}[1]{\texttt{#1}}
\expandafter\ifx\csname urlstyle\endcsname\relax
  \providecommand{\doi}[1]{doi: #1}\else
  \providecommand{\doi}{doi: \begingroup \urlstyle{rm}\Url}\fi

\bibitem[Abramitzky et~al.(2012)Abramitzky, Einav, Kolkowitz, and
  Mill]{abramitzky2012linecalls}
Ran Abramitzky, Liran Einav, Shimon Kolkowitz, and Roy Mill.
\newblock On the optimality of line call challenges in professional tennis.
\newblock \emph{International Economic Review}, 53\penalty0 (3):\penalty0
  939--964, 2012.
\newblock \doi{10.1111/j.1468-2354.2012.00706.x}.
\newblock URL \url{https://doi.org/10.1111/j.1468-2354.2012.00706.x}.

\bibitem[Deshpande and Wyner(2017)]{deshpande2017pitchframing}
Sameer~K. Deshpande and Abraham~J. Wyner.
\newblock A hierarchical {Bayesian} model of pitch framing.
\newblock \emph{Journal of Quantitative Analysis in Sports}, 13\penalty0
  (3):\penalty0 95--112, 2017.
\newblock \doi{10.1515/jqas-2017-0027}.
\newblock URL \url{https://doi.org/10.1515/jqas-2017-0027}.

\bibitem[Efron and Tibshirani(1994)]{efron1994bootstrap}
Bradley Efron and Robert~J. Tibshirani.
\newblock \emph{An Introduction to the Bootstrap}.
\newblock Chapman \& Hall, 1994.
\newblock URL
  \url{https://www.routledge.com/An-Introduction-to-the-Bootstrap/Efron-Tibshirani/p/book/9780412042317}.

\bibitem[Gelman et~al.(2013)Gelman, Carlin, Stern, Dunson, Vehtari, and
  Rubin]{gelman2013bayesian}
Andrew Gelman, John~B. Carlin, Hal~S. Stern, David~B. Dunson, Aki Vehtari, and
  Donald~B. Rubin.
\newblock \emph{{Bayesian} Data Analysis}.
\newblock Chapman \& Hall/CRC, 3 edition, 2013.
\newblock URL
  \url{https://www.routledge.com/Bayesian-Data-Analysis/Gelman-Carlin-Stern-Dunson-Vehtari-Rubin/p/book/9781439840955}.

\bibitem[Gneiting and Raftery(2007)]{gneiting2007scoring}
Tilmann Gneiting and Adrian~E. Raftery.
\newblock Strictly proper scoring rules, prediction, and estimation.
\newblock \emph{Journal of the American Statistical Association}, 102\penalty0
  (477):\penalty0 359--378, 2007.
\newblock \doi{10.1198/016214506000001437}.
\newblock URL \url{https://doi.org/10.1198/016214506000001437}.

\bibitem[Lee et~al.(2026)Lee, Han, Lee, and Ko]{lee2026algorithmic}
Kichang Lee, Gyeongmin Han, Sungmin Lee, and JeongGil Ko.
\newblock When the strike zone becomes algorithmic: Umpire judgment and player
  challenge decisions under {AI} review.
\newblock arXiv preprint arXiv:2609.25525, 2026.
\newblock URL \url{https://arxiv.org/abs/2609.25525}.

\bibitem[Lindsey(1963)]{lindsey1963strategies}
George~R. Lindsey.
\newblock An investigation of strategies in baseball.
\newblock \emph{Operations Research}, 11\penalty0 (4):\penalty0 477--501, 1963.
\newblock \doi{10.1287/opre.11.4.477}.
\newblock URL \url{https://doi.org/10.1287/opre.11.4.477}.

\bibitem[{Major League Baseball}(2025)]{mlb2025absrules}
{Major League Baseball}.
\newblock {MLB} announces {ABS} challenge system coming to the major leagues
  beginning in the 2026 season.
\newblock Major League Baseball press release, 2025.
\newblock URL
  \url{https://www.mlb.com/press-release/press-release-mlb-announces-abs-challenge-system-coming-to-the-major-leagues-beginning-in-the-2026-season}.
\newblock Published September 23, 2025; accessed October 21, 2026.

\bibitem[{Major League Baseball}(2026{\natexlab{a}})]{mlb2026abs}
{Major League Baseball}.
\newblock {ABS} challenge dashboard.
\newblock Baseball Savant, 2026{\natexlab{a}}.
\newblock URL \url{https://baseballsavant.mlb.com/abs}.
\newblock Accessed October 17, 2026.

\bibitem[{Major League Baseball}(2026{\natexlab{b}})]{mlb2026savant}
{Major League Baseball}.
\newblock Statcast search {CSV} documentation.
\newblock Baseball Savant, 2026{\natexlab{b}}.
\newblock URL \url{https://baseballsavant.mlb.com/csv-docs}.
\newblock Accessed October 17, 2026.

\bibitem[{Major League Baseball}(2026{\natexlab{c}})]{mlb2026statsapi}
{Major League Baseball}.
\newblock {MLB Stats API} documentation, 2026{\natexlab{c}}.
\newblock URL \url{https://docs.statsapi.mlb.com/}.
\newblock Accessed October 17, 2026.

\bibitem[Martinez(2026)]{martinez2026abs}
Christopher Martinez.
\newblock When to challenge a pitch? a reinforcement learning approach to {ABS}
  challenge strategy.
\newblock Research presentation, SABR Analytics Conference, 2026.
\newblock URL \url{https://sabr.org/analytics/2026-photos-2}.
\newblock February 28, 2026.

\bibitem[Puterman(1994)]{puterman1994mdp}
Martin~L. Puterman.
\newblock \emph{Markov Decision Processes: Discrete Stochastic Dynamic
  Programming}.
\newblock John Wiley \& Sons, 1994.
\newblock \doi{10.1002/9780470316887}.
\newblock URL \url{https://doi.org/10.1002/9780470316887}.

\bibitem[Wood(2017)]{wood2017gam}
Simon~N. Wood.
\newblock \emph{Generalized Additive Models: An Introduction with {R}}.
\newblock Chapman \& Hall/CRC, 2 edition, 2017.
\newblock URL
  \url{https://www.routledge.com/Generalized-Additive-Models-An-Introduction-with-R-Second-Edition/Wood/p/book/9781315370279}.

\bibitem[Yee and Deshpande(2024)]{yee2024platediscipline}
Ryan Yee and Sameer~K. Deshpande.
\newblock Evaluating plate discipline in {Major League Baseball} with {Bayesian
  Additive Regression Trees}.
\newblock \emph{Journal of Quantitative Analysis in Sports}, 20\penalty0
  (1):\penalty0 5--20, 2024.
\newblock \doi{10.1515/jqas-2023-0048}.
\newblock URL \url{https://doi.org/10.1515/jqas-2023-0048}.

\end{thebibliography}

\clearpage
\appendix
\begin{center}
\Large\bfseries Appendix
\end{center}
\setcounter{equation}{0}
\renewcommand{\theequation}{A.\arabic{equation}}
\begingroup

\section{Estimation of run values and future opportunities}
\label{app:estimation}

The policy in Section~\ref{sec:policy} combines estimates of runs remaining, pitch-location uncertainty, and the opportunities a team may encounter later in the game.
The run-expectancy table uses 2023--2025 data, while the remaining estimates use only the 1,442 development games defined in Section~\ref{sec:samples}.
The calculations below describe the primary policy, which assigns the fitted error widths entirely to sensory uncertainty.

\subsection{Estimating runs remaining}
\label{app:run-expectancy}

Run expectancy is the expected number of runs the batting team will score from a pitch through the end of the half-inning.
The 2023--2025 Statcast records provide one observation per pitch, including pitches that were swung at.
The final batting-team score for each half-inning is reconstructed as the largest score recorded immediately before or after any pitch in that half-inning.
For pitch $j$, the response $Y_j$ is this final score minus the batting-team score before the pitch.
It therefore includes any runs scored on that pitch and all later runs in the same half-inning.
Repeated observations of a count, such as pitches following a two-strike foul, remain separate observations.

The resulting table contains $8\times3\times12=288$ base-out-count states.
The shrinkage target is the average number of runs remaining among all pitches with the same base occupancy and number of outs, pooling over counts.
In Equation~\eqref{eq:re-shrinkage}, $\lambda_{\mathrm{RE}}$ is the number of additional pitches represented by that pooled mean.
A state with no observations receives the pooled base-out mean; if that base-out state is also unobserved, we use the average number of runs remaining in the historical sample.
The table therefore provides an estimate for every base-out-count state, with zero runs remaining once the half-inning has ended.

Predictions for later seasons determine how strongly the estimates borrow across counts, as described in Appendix~\ref{app:pooling-validation}.
The selected strength is then used to estimate run expectancy from all three historical seasons, and this table remains fixed throughout development and evaluation of the 2026 policy.

\subsection{Reconstructing pitch margins}
\label{app:geometry}

We reconstruct the strike zone as a 17-inch-wide rectangle centered over the plate.
Its upper and lower limits come from the live feed, with Statcast limits used when the corresponding feed value is missing.
We round these limits to three decimal places in feet before converting them to inches.
Pitch-center coordinates come from Statcast, using only records matched to the same batter and pitcher in the game feed.

Let $x$ be the horizontal displacement of the pitch center from the center of home plate, and let $z$ be its height.
The lower and upper zone limits, $\ell$ and $u$, share the same vertical reference as $z$, and all distances are measured in inches.
Set $q_x=|x|-8.5$ and $q_z=\max(\ell-z,z-u)$.
With an assumed ball radius of 1.45 inches, the signed distance from the ball to the zone is
\begin{equation}
 D=\sqrt{\max(q_x,0)^2+\max(q_z,0)^2}
   +\min\{\max(q_x,q_z),0\}-1.45.
 \label{eq:appendix-geometry}
\end{equation}
The square-root term measures distance to the nearest edge or corner when the pitch center lies outside the rectangle.
Inside the rectangle, the min--max term gives the negative distance to the nearest edge.
Subtracting the ball's radius accounts for pitches whose centers miss the rectangle but whose edges touch it.
The reconstructed ruling is a strike when $D\leq0$ and a ball otherwise.

We orient the margin toward the potential challenger, taking $M=D$ on offense and $M=-D$ on defense.
A called strike is correctable when $M>0$; a called ball is correctable when $M\geq0$, because contact with the boundary counts as a strike.

\subsection{Estimating the margin priors}
\label{app:margin-priors}

The prior for the signed pitch margin $M$ describes where a pitch is likely to lie relative to the strike-zone boundary before the player's impression is taken into account.
It conditions only on the umpire's call and the pre-pitch count, with called strikes informing the offensive prior and called balls informing the defensive prior.
To represent the opportunities players face, these distributions draw on every eligible tracked call in the development sample, including those left unchallenged and those made after a team had exhausted its challenges.
The margins are oriented toward the potential challenger as in Section~\ref{sec:observed-and-perceived-location} and reconstructed from pitch tracking rather than inferred from review outcomes.

Margins are grouped into bins of width $h=0.01$ inches, so an observed margin $M$ belongs to bin $b=\lfloor M/h\rfloor$, with center $m_b=(b+1/2)h$.
Each role's prior covers its full observed range, including pitches beyond the three-inch window used to calibrate error widths (Appendix~\ref{app:error-calibration}), and assigns zero mass to bins without observations for that role.

For role $r$, the role-wide bin mass is $p_r(b)=n_{rb}/n_r$, where $n_{rb}$ counts observations in bin $b$ and $n_r$ counts all observations for that role.
Under Equation~\eqref{eq:margin-prior}, count-specific estimates borrow from this role-wide distribution, allowing a bin to receive positive mass even when it is unobserved at that count.
If a count does not appear in the games used to estimate the prior, its prior is the role-wide distribution.
In the posterior and validation calculations, each bin's probability is concentrated at its center.

To judge how well a prior describes new pitches, we compare each held-out pitch's tracked margin with the distribution estimated from other development games.
For the original tracked margin $y$ and discrete predictive distribution $p$, the continuous ranked probability score (CRPS) is
\begin{equation}
 \mathrm{CRPS}(p,y)
 =\sum_b p(b)|m_b-y|
 -\frac12\sum_b\sum_{b'}p(b)p(b')|m_b-m_{b'}|.
 \label{eq:prior-crps}
\end{equation}
Smaller scores indicate better predictive distributions \citep{gneiting2007scoring}.

The game-based validation in Appendix~\ref{app:pooling-validation} determines each role's pooling strength $\alpha_r$ by mean CRPS.
At the selected strength, the prior draws on all development games and remains fixed throughout policy construction and test-game evaluation.

\subsection{Calibrating the error widths}
\label{app:error-calibration}

The error widths describe how sharply challenge probability changes with tracked margin after accounting for the circumstances of the call.
We estimate this relationship from challenged and unchallenged pitches in the development games, including calls with nonpositive estimated correction gain.
The sample is restricted to eligible calls within three inches of the boundary for which the team had at least one challenge available.

For role $r$, a probit model accounts for the circumstances of each decision through
\begin{equation}
\begin{aligned}
 \Phi^{-1}\P(A_i^{\mathrm{hist}}=1\mid M_i,X_i,r)
 ={}&\beta_{0r}+b_rM_i
     +f_{Gr}(G_i)+f_{Ir}(I_i)+f_{dr}(d_i)\\
    &+\boldsymbol Z_i^\top\boldsymbol\gamma_r
     +u_{a(i),r}+v_{T(i),r}.
\end{aligned}
 \label{eq:appendix-calibration}
\end{equation}
Here $I_i$ is the inning and $d_i$ is the potential challenging team's score minus its opponent's score.
The vector $\boldsymbol Z_i$ represents categorical effects for the pre-pitch count, outs, remaining challenges, pitch family, and batter--pitcher handedness combination.

We group four-seam fastballs, generic fastballs, sinkers, and cutters as fastballs; sliders, sweepers, slurves, curves, knuckle curves, and slow curves as breaking balls; and changeups, splitters, forkballs, and screwballs as offspeed pitches.
All remaining or unidentified types form a separate category.

Participant and team intercepts allow willingness to challenge to vary across players and teams.
The participant intercept $u_{a(i),r}$ applies to the batter on offense and the pitcher--catcher pair on defense, while the team intercept $v_{T(i),r}$ applies to the batting and fielding teams, respectively.
These effects are independent, mean-zero Gaussians, with separate participant and team variances within each role.

Correction gain, inning, and score margin can affect willingness to challenge nonlinearly, so their effects are represented by cubic regression splines with basis dimensions 7, 7, and 9, respectively.
Smoothing parameters and random-effect variances are estimated by restricted maximum likelihood \citep{wood2017gam}.
Margin enters linearly, so its fitted coefficient gives the combined error width $w_r=1/b_r$, while the other terms account for contextual differences in willingness to challenge.
Calibration relies on whether players challenged each call, without using the review outcome, and the resulting widths remain fixed throughout policy construction and test-game evaluation.

\subsection{Estimating the sequence of future opportunities}
\label{app:future-opportunities}

To value a challenge saved for later, we estimate at what stage of the game the team's next called pitch may occur and the chance that no further call remains.
Following these possibilities from one call to the next describes the opportunities the team may encounter during the rest of the game.

In each development game, we follow the team's called strikes on offense and called balls on defense.
Each call is paired with the next call in that team's sequence, while the final call records that none follows.
Calls without ABS eligibility or usable tracking remain as mandatory passes, and calls made after a team has exhausted its challenges remain because a different strategy might have preserved one.
The resulting record contains 212,420 called pitches, including the 211,596 eligible tracked pitches used to construct potential challenges.

The state $z$ records the inning, top or bottom half, outs, and the team's role $r$, either offense or defense.
Regulation innings remain distinct, while corresponding situations in all extra innings share a state.
We assume that $z$ determines the distribution of subsequent calls, averaging over differences in the current count, correction value, and perceived margin.
This distribution is unchanged by the current challenge decision.

Calls observed in state $z$ form the sample $\mathcal I_z$, containing $n_z=|\mathcal I_z|$ development observations.
All 120 states are represented.
To assess whether information from other innings improves these estimates, we also consider a broader group of comparable states.
We denote this group by $b(z)$: states with the same role, top/bottom half, and number of outs as $z$, pooled across innings while keeping regulation and extra innings separate.
The associated sample $\mathcal I_{b(z)}$ contains all calls from these states, including those in $\mathcal I_z$.

Borrowing information from another inning requires preserving how play progressed between calls.
For example, a visiting team's sequence from one out in the top of the fourth to no outs in the bottom of the fourth can inform the same progression in the seventh: two outs pass, and the team switches from batting to fielding.
More generally, we advance the current game situation by the observed number of completed outs, counting outs for both teams, and retain the team's role at the next call.
An observed end to the sequence still contributes no further call.

The progression between calls also matters when it carries the team into a later extra inning, restoring a challenge if none remain.
Each continuing outcome therefore records both the next state $z'$ and an indicator $e$ of entry to a later extra inning, with $e=1$ for entry and $e=0$ otherwise.
For both observed and borrowed progressions, entry is determined before grouping extra innings into shared states.

These recorded outcomes give two empirical distributions: $\widehat P_z^{\mathrm{state}}$ from calls in $\mathcal I_z$, and $\widehat P_z^{\mathrm{pool}}$ from the broader pool after translating its progressions to the current state.
Both include the outcome that no further call occurs, with probabilities calculated over all calls in the corresponding sample.
Their mixture estimates what follows a call in state $z$:
\begin{equation}
 \widehat P_z
 =\omega_z\widehat P_z^{\mathrm{state}}
  +(1-\omega_z)\widehat P_z^{\mathrm{pool}},
 \qquad
 \omega_z=\frac{n_z}{n_z+\lambda_F}.
 \label{eq:appendix-state-weight}
\end{equation}
The broader pool contributes the equivalent of $\lambda_F$ calls, giving it greater influence in states with fewer observations.
The same strength governs the count and correction-gain distributions in Appendix~\ref{app:opportunity-expectation}.
The selected $\lambda_F=0$ gives $\omega_z=1$, so the primary transition estimates use $\widehat P_z^{\mathrm{state}}$.
Appendix~\ref{app:pooling-validation} describes this selection and the sensitivity comparisons with positive pooling strengths.

Together with the challenge-restoration rule $g(k,e)$, this distribution determines the value of later opportunities through Equation~\eqref{eq:appendix-continuation}.

\subsection{Averaging over future counts, gains, and impressions}
\label{app:opportunity-expectation}

With a challenge available, a team chooses between requesting a review and preserving its challenges for later.
The value of that choice depends on the count, the runs an overturn could recover, and the player's impression of the pitch.
Averaging the net benefit of the better decision, measured relative to passing, gives the term added to $C_k(z)$ to obtain $V_k(z)$ in Equation~\eqref{eq:bellman}.

The calls observed in each state and the broader pool of comparable calls describe which counts and correction gains occur together, including calls for which the policy must pass.
Within these samples, $a_i=1$ denotes a call that can be evaluated as a challenge and has $G_i>0$, while $a_i=0$ denotes a call requiring a pass.
Let $\widehat H_z$ be the empirical distribution of $(c_i,G_i,a_i)$ in $\mathcal I_z$.
The corresponding distribution $\widehat H_{b(z)}$ uses the pooled calls from $b(z)$, the broader group of comparable states defined in Appendix~\ref{app:future-opportunities}.
The distribution used in the Bellman expectation is
\begin{equation}
 \widetilde H_z=\omega_z\widehat H_z+(1-\omega_z)\widehat H_{b(z)}.
 \label{eq:appendix-opportunity-distribution}
\end{equation}
This mixture preserves how often different counts and correction gains occur together, without imposing a separate parametric distribution on $G$.
At the selected $\lambda_F=0$, it reduces to the empirical distribution $\widehat H_z$ within the state.

For a role $r$ and count $c$, the prior $p_{rc}(m)$ and sensory model imply the impression density
\begin{equation}
 f_{rc}(s)=\sum_m p_{rc}(m)\phi_{\sigma_{s,r}}(s-m).
 \label{eq:appendix-signal-density}
\end{equation}
Each possible margin contributes a normal distribution of impressions, weighted by the prior probability of that margin.

The value calculation averages first over the player's impressions for a given count and correction gain, and then over the counts and gains encountered in state $z$.
At each impression, challenging contributes its net gain if that is positive, while passing adds nothing to the continuation value.
Using $q_{rc}(s)$ as shorthand for $q_{rc}(s;\sigma_{s,r})$, the expected net benefit for a team with $k\geq1$ challenges is
\begin{equation}
 \int_{\{a=1\}}\left\{\int
 \big[q_{rc}(s)G-\{1-q_{rc}(s)\}L_k(z)\big]_+
 f_{rc}(s)\,ds\right\}\,d\widetilde H_z(c,G,a),
 \label{eq:appendix-bellman-expectation}
\end{equation}
where $[x]_+=\max(0,x)$ and $r$ is the role in $z$.
Calls requiring a pass remain in the distribution and transition record but contribute zero to the expectation.
With no challenge available at the current call, the team must pass, so the expectation is zero and $V_0(z)=C_0(z)$.

The same inner average can be calculated from the probabilities of successful and unsuccessful challenges.
With $L=L_k(z)$, a call with $G>0$ and $L>0$ is challenged when the impression exceeds the cutoff $t$ satisfying $q_{rc}(t)=L/(G+L)$.
The probabilities of challenging, $A_{rc}(t)$, and of both challenging and overturning the call, $B_{rc}(t)$, are
\begin{align}
 A_{rc}(t)&=\sum_m p_{rc}(m)\,
 \overline\Phi\!\left(\frac{t-m}{\sigma_{s,r}}\right),\notag\\
 B_{rc}(t)&=\sum_{m>0} p_{rc}(m)\,
 \overline\Phi\!\left(\frac{t-m}{\sigma_{s,r}}\right),
 \label{eq:appendix-cutoff-probabilities}
\end{align}
where $\overline\Phi=1-\Phi$.
The difference $A_{rc}(t)-B_{rc}(t)$ is the probability of an unsuccessful challenge.
The expected net gain $GB_{rc}(t)-L\{A_{rc}(t)-B_{rc}(t)\}$ therefore subtracts the expected future cost of failed reviews from the expected correction value of successful ones.
When $G>0$ and $L=0$, a failed review has no future cost, so $t=-\infty$.

Numerical evaluation of $f_{rc}$ and $q_{rc}$ uses a 0.005-inch grid covering $[-40,40]$ inches and extending at least $12\sigma_{s,r}$ beyond the smallest and largest margin centers.
Gaussian-mixture tails and trapezoidal integration give $A_{rc}$ and $B_{rc}$, with cutoffs in perceived margin interpolated from posterior log odds.

Because calls can recur within a game stage and extra innings reuse the same states, $C_k(z)$ and $V_k(z)$ must be solved together.
Starting with $V_k(z)=0$, value iteration repeatedly updates the value of later opportunities, $C_k(z)$, and the value including the current opportunity, $V_k(z)$, until no $V_k(z)$ changes by more than $10^{-7}$ runs.

\subsection{Choosing pooling strengths and assessing sensitivity}
\label{app:pooling-validation}

We choose pooling strengths by their performance on held-out historical or development observations, reserving the 983 later games for policy evaluation.
Run expectancy ($\lambda_{\mathrm{RE}}$), margin priors ($\alpha_r$), and future opportunities ($\lambda_F$) share the candidate set
\[
 \mathcal A=\{0,1,3,10,30,100,300,1000,3000,10000\}.
\]
The positive candidates cover approximately logarithmic intervals, and zero leaves an observed state or count unpooled.

For run expectancy, the 2023 observations predict runs remaining in 2024, and the 2023--2024 observations predict those in 2025.
The selected strength minimizes mean squared error across both validation seasons.
For margin priors, each of five approximately equal folds of whole development games is predicted from the other four, and the selected strength for each role minimizes mean CRPS (Equation~\eqref{eq:prior-crps}).
Both criteria give each pitch equal weight, and all distributions used for pooling are estimated from the training observations alone.

For future opportunities, the same five folds compare how much correction value the resulting challenge policies recover.
The selected run-expectancy and margin-prior strengths remain fixed (Table~\ref{tab:pooling-validation}).
For each held-out fold, the other four supply the fitted priors, error widths, and future-opportunity distributions, while run expectancy comes from all 2023--2025 observations.
Candidate policies within a fold therefore share the fitted perception model and correction values, differing only in how much they pool future opportunities.

Each policy is evaluated along the held-out recorded opportunities using the procedure in Section~\ref{sec:whole-game-evaluation}.
The selected strength maximizes mean modeled correction runs recovered by both teams, giving each of the 1,442 development games equal weight.
This selects $\lambda_F=0$, although the criterion is nearly flat near zero: $\lambda_F=30$ reduces total modeled recovery by only 0.05 runs across these games.
The selected models are then fitted to the full historical and development samples for evaluation on the later games.

\begin{table}[htbp]
\centering
\caption{Pooling strengths selected from historical and development observations.}
\label{tab:pooling-validation}
\small
\renewcommand{\arraystretch}{1.1}
\begin{tabular*}{\linewidth}{@{\extracolsep{\fill}}llr@{}}
\toprule
Component & Validation criterion & Selected weight \\
\midrule
Run expectancy, $\lambda_{\mathrm{RE}}$ & Squared error & 100 \\
Offensive margin prior, $\alpha_o$ & CRPS & 30 \\
Defensive margin prior, $\alpha_d$ & CRPS & 10 \\
Future opportunities, $\lambda_F$ & Mean game value & 0 \\
\bottomrule
\end{tabular*}
\end{table}

The policy's estimated gain changes little across nearby pooling strengths (Table~\ref{tab:pooling-sensitivity}).
The gains range from 0.0828 to 0.0839 modeled correction runs per game over observed decisions, compared with 0.0839 for the selected policy.

These comparisons change one weight at a time to neighboring grid values, keeping the other weights at their selected values.
Because $\lambda_F=0$ lies at the lower boundary, its alternatives are the two smallest positive candidates, 1 and 3.
All policies are evaluated with the selected run-expectancy table ($\lambda_{\mathrm{RE}}=100$), so the same overturned call has the same value in every comparison.

\begin{table}[htbp]
\centering
\caption{Policy gains at nearby pooling strengths, measured in modeled correction runs per game over observed decisions.}
\label{tab:pooling-sensitivity}
\small
\renewcommand{\arraystretch}{1.1}
\begin{tabular*}{\linewidth}{@{\extracolsep{\fill}}lrr@{}}
\toprule
Pooling weight & Alternative & Gain (runs/game) \\
\midrule
Run expectancy, $\lambda_{\mathrm{RE}}$ & 30 & 0.0838 \\
 & 300 & 0.0828 \\
\addlinespace
Offensive margin prior, $\alpha_o$ & 10 & 0.0839 \\
 & 100 & 0.0839 \\
\addlinespace
Defensive margin prior, $\alpha_d$ & 3 & 0.0839 \\
 & 30 & 0.0839 \\
\addlinespace
Future opportunities, $\lambda_F$ & 1 & 0.0839 \\
 & 3 & 0.0839 \\
\bottomrule
\end{tabular*}
\end{table}

\endgroup

\end{document}